\documentclass[printer]{aa} % for a referee version

\usepackage{graphicx}
\usepackage{natbib}
\usepackage{txfonts}
\usepackage{url}
\usepackage{multirow}
\usepackage{verbatim}
\begin{document}

\title{Properties of simulated Milky Way-mass galaxies in loose group and field environments}
\author{C.~G. Few \inst{1}
        \and B.~K. Gibson \inst{1,2,3}
        \and S. Courty\inst{4}
        \and L. Michel-Dansac \inst{4}
        \and C.~B. Brook \inst{5}
        \and  G.~S. Stinson \inst{6}
}       
\institute{Jeremiah Horrocks Institute, University of Central Lancashire, Preston, PR1~2HE, UK\\
  \email{cgfew@uclan.ac.uk}
  \and Department of Astronomy \& Physics, Saint Mary's University, Halifax, Nova Scotia, B3H 3C3, Canada
  \and Monash Centre for Astrophysics, Monash University, Victoria, 3800, Australia
  \and Universit\'e de Lyon; Universit\'e Lyon~1, Observatoire de Lyon, 9 avenue Charles Andr\'e, Saint-Genis Laval, F-69230, France;
       CNRS, UMR 5574, Centre de Recherche Astrophysique de Lyon; Ecole Normale Sup\'erieure de Lyon
  \and  Grupo de Astrof\'isica, Departamento de Fisica Teorica, Modulo C-15, Universidad Aut\'onoma de Madrid, 28049 Cantoblanco, Spain
  \and Max-Planck-Institut f\"{u}r Astronomie, K\"{o}nigstuhl 17, 69117 Heidelberg, Germany
}

\date{Received May 22, 2012}
%__________________________________________________________________
\abstract
{}
{We test the validity of comparing simulated field disk galaxies with the empirical properties of 
systems situated within environments more comparable to loose groups, including the Milky Way's Local Group.}
{Cosmological simulations of Milky Way-mass galaxies have been realised in two different environment samples: in the field and 
in loose groups environments with similar properties to the Local Group. Apart from the differing environments of the galaxies, the samples are 
kept as homogeneous as possible with equivalent ranges in last major merger time, halo mass and halo spin. Comparison of these 
two samples allow for systematic differences in the simulations to be identified. A kinematic decomposition is employed 
to objectively quantify the spheroid-to-disk ratio and to isolate the disk-star population. Metallicity gradients, disk scale lengths, 
colours, magnitudes and age-velocity dispersion relations are studied for each galaxy in the suite and the strength of the link between these and 
environment of the galaxies is studied.}
{Metallicity gradients are consistent with observations of HII regions in spiral galaxies and, in agreement with observations, 
correlate with total galaxy mass. The bulge-to-disk ratio of the galaxies show that these galaxies are less spheroid dominated than many other simulated galaxies 
in literature with the majority of both samples being disk dominated. We find that secular evolution and mergers dominate the spread of morphologies and metallicity 
gradients with no visible differences between the two environment samples. In contrast with this consistency in the two samples there is tentative evidence for a 
systematic difference in the velocity dispersion-age relations of galaxies in the different environments. Loose group galaxies appear to have more discrete steps 
in their velocity dispersion-age relations, if this is true it suggests that impulsive heating is more efficient in the stars of galaxies in denser environment than 
in the field. We conclude that at the current resolution of cosmological galaxy simulations field environment galaxies are sufficiently similar to those 
in loose groups to be acceptable proxies for comparison with the Milky Way provided that a similar assembly history is considered.
}
{}

%Add keywords
\keywords{Galaxies: Local Group, formation, evolution -- methods: numerical}

\titlerunning{Properties of simulated Milky Way-mass galaxies}
\authorrunning{C. G. Few et al.}

\maketitle

%__________________________________________________________________
\section{Introduction}
\label{intro}

It is well established that galaxy interactions and mergers result in significant changes in
a system's star formation rate \citep{barton00,lambas03,nikolic04,ellison08} and its chemical properties 
\citep{donzelli00,marquez02,fabbiano04,kewley06,micheldansac08,rupke08,kewley10, solalonso10}. Studies have 
shown that in denser large scale environments the star formation rate of galaxies tends to be 
lower \citep{gomez03}. It is often suggested that galaxies in clusters are older and have therefore consumed the limited gas 
available for star formation \citep{lilly96}, or that proximity to other galaxies means that the reservoir 
of infalling gas must be shared \citep{Lewis02}. It is also thought that a dominant method of reducing the 
star formation rate in dense environment is through ram pressure stripping that removes the gas envelope from 
field galaxies as they are accreted to groups and clusters \citep{balogh04b}. 

Proximity to a cluster centre is also known to impact the morphology of galaxies and clusters have a greater fraction 
of early type galaxies; i.e. the so-called morphology-density relation \citep{dressler97}. While the accretion of field galaxies into denser environments may strip 
gas from the galaxy and leave a slowly reddening S0 galaxy, it will do little else to alter the morphology. Morphological 
transformations are therefore attributed to gravitational interactions with other group members \citep{moore96, weinmann06}. 
Denser environments also increase the likelihood of galaxy mergers. This is supported by the findings of \cite{mcgee08} 
where an enhancement in the number of asymmetric disks is found in group environments (these groups typically have velocity 
dispersions less than 700~km~s$^{-1}$, smaller than the larger clusters considered in many other works). The authors also put forward 
the intriguing conclusion that while the groups exhibit a larger fraction of galaxies that are bulge dominated, 
\emph{no evidence is found that the group environment has any effect on the bulk properties of the disk galaxy population}.

Mergers have a direct impact on both the star formation history of the galaxies and the metal distribution. Hydrodynamical 
simulations by \cite{hernquist89} and \cite{barnes96} show that mergers funnel gas into the central regions of galaxies. 
This would tend to dilute the gas phase metallicity at small radii and trigger centrally concentrated star formation 
\citep{rupke10, montuori10,perez11}. This trend is consistent with observations of flattened metallicity gradients in luminous and 
ultraluminous infrared galaxies that are identified as merged systems \citep{rupke08} and in interacting galaxy pairs 
\citep{ellison08, micheldansac08, kewley10} where the influence on star formation rates extends to projected separations 
of up to 40~h$^{-1}$~kpc \citep{ellison08}. 

Interactions are far more common in denser environments and one would expect to see the effect of these 
interactions imprinted on the metallicity of cluster and group galaxies when compared with those in the field. 
\cite{cooper08} observe that not only do members of clusters have greater metallicities but also that on average 
galaxies that are closer to other cluster members have greater oxygen abundances (an effect of order 0.05 dex). The conclusion 
of the work is that metallicity effects are not driven by the cluster as a whole but only by the specific proximity 
of each galaxy to others; consistent with past findings \citep{balogh04, martinez08}. 
The independence of metallicity on large-scale environment is perhaps refuted by the findings of \cite{ellison09} where a residual 
metallicity-environment effect is found observationally even after the dependence on luminosity and colour have been accounted for, however these 
two results are not entirely irreconcilable. \cite{martinezvaquero09} select simulated dark matter systems based on the 
mass and circular velocity of haloes, the mutual proximity of the haloes and the distance to a halo with the
same mass as the Virgo cluster. These simulations explore the properties of the haloes while relaxing these 
criteria and find that the nearness of massive external haloes is the most significant factor determining the dispersion 
of Local Group systems: the coldness of the Local Group can be attributed mostly to its isolation from clusters.

Recent simulations have employed higher resolution and more advanced supernova feedback 
prescriptions which ameliorate the traditional failings of galaxy simulations \citep{robertson04,governato07,scannapieco09,
sanchezblazquez09,stinson10,rahimi10,brooks11}. It is now possible (and prudent) 
to examine the more subtle factors that influence galaxy properties. Given the demonstrable difference 
between galaxies in different (albeit extremely so) environments it is reasonable to expect that galaxies 
in loose groups such as our own Local Group may differ from those in field.

At present the majority of literature on this topic focusses on constrained simulations based upon cosmological initial conditions that will 
purposefully give rise to systems with the properties of the Local Group \citep{gottloeber10, peirani10, libeskind11, peirani12} or the use of extremely large simulation 
volumes to search for analogous systems \citep[][and references therein]{snaith11}. Here we provide a complementary 
approach by identifying Local Group analogues in a suite of hydrodynamical simulations.  In what follows, we will 
explore the hypothesis that simulated isolated field galaxies can be considered suitable proxies for Local Group analogues.

The purpose of this work is not to reproduce artificial clones of the Local Group but rather to gauge the systematic offset in 
properties between field galaxies and those with a similar degree of interaction with neighbours such as is encountered between the Milky Way and 
Andromeda. These  groups are henceforth termed as ``loose groups'' to be clear that the systems in the loose group sample are not 
Local Group clones from initial conditions designed to reproduce the exact layout of the local universe but are chosen from cosmological 
simulations based on certain similarities to the Local Group. In doing so we find galaxies with a range of merger histories that nonetheless 
produce disk galaxies. It is hoped that the discrepancy between the properties of these two samples will 
place constraints on future simulations and present insight into the failings of field galaxy simulations when 
attempting to recover the properties of Local Group galaxies. The method employed in producing this sample and the properties 
of the simulation code are described in \S\ref{meth}. In \S\ref{resu} we present an analysis of the galaxy 
disk fractions, colours, magnitudes, metallicity gradients and velocity dispersions and discuss these results in \S\ref{disc}.

%__________________________________________________________________
\section{Method}
\label{meth}
%__________________________________________________________________
\subsection{Simulations}
\label{sims}

The galaxies presented here were simulated using the adaptive mesh
refinement code \textsc{ramses} (v3.01)
\citep{teyssier02}. \textsc{ramses} is a three-dimensional Eulerian
hydrodynamical code with an N-body particle-mesh scheme to compute
self-gravity. The mesh automatically refines according to the local
particle density in addition to a static refinement of nested regions
that reduces the run-time while maintaining high resolution around the
galaxy of interest. Details of the refinement scheme are described by
\cite{teyssier02}. \textsc{ramses} includes density- and
metallicity-dependent radiative cooling rates, using an ionisation
equilibrium with an ultra-violet radiative background
\citep{haardt96}. Gas cells with a density $\rho$ exceeding a star 
formation threshold of $\rho_0$ = 0.1~cm$^{-3}$ form stars at a rate of 
$\dot{\rho} = -\rho/t_{\star}$. The star formation timescale $t_{\star}$ 
is itself a function of the density and the free-fall time through the 
free parameter, $t_0$, as follows: $t_{\star} = t_0(\rho/\rho_0$)$^{-1/2}$ \citep{dubois08}. 
We use $t_0=8$ Gyr, corresponding to a 2$\%$ star formation efficiency for $\rho_0$ = 0.1~cm$^{-3}$. 
We use the kinetic feedback mode
of \textsc{ramses} that aims to reproduce the energetic and chemical
enrichment of SNeII explosions: After $10^7$ years, star particles
release some mass, momentum and energy into a 2-cell radius
feedback-sphere centred on the star particle. We characterise the initial 
mass fraction of the star particles with the parameter $\eta_{\mathrm{SN}}=10\%$ (which 
corresponds to the mass fraction of stars contributing supernovae feedback) 
and we do not use any mass loading factor. The energy injected into the gas 
phase is in the form of kinetic energy with 100\% efficiency (i.e. 10$^{51}$ erg
SN$^{-1}$). Chemical enrichment of the gas is followed through the
global metallicity Z using a yield of 10$\%$. \textsc{ramses} includes a polytropic 
equation of state with an index of $5/3$ to prevent non-physical gas fragmentation 
in all gas cells with an hydrogen density larger than the star formation 
density threshold. The temperature threshold is $T_{\mathrm{th}}=10^4$~K and in the 
analysis and Table~\ref{tab2} in particular, we will set the temperature of 
the high-density gas cells to $10^4$ K to account for unresolved cold gas. 
A detailed description of the star formation and feedback treatments may be found in
\cite{dubois08}, though the v3.01 \textsc{ramses} has a different
technical implementation of the kinetic feedback.

Candidate haloes for this work were identified from dark matter
simulations and then resimulated with baryonic physics and a more
refined grid around regions of interest, using the same technique as
in \cite{sanchezblazquez09}. The simulations are conducted in a
cosmological framework with $H_0$=70~km s$^{-1}$Mpc$^{-1}$,
$\Omega_{\mathrm{m}}$=0.28, $\Omega_{\mathrm{\Lambda}}$=0.72, $\Omega_{\mathrm{b}}$=0.045, and
$\sigma_8$=0.8. Two volumes are used with a size of 20~$h^{-1}$~Mpc
and 24~$h^{-1}$~Mpc and the maximum refinement achieved (16 levels)
corresponds to 436~pc and 523~pc respectively. The dark matter mass
resolution is 5.5$\times$10$^{6}$~M$_\odot$ and 9.5$\times$10$^{6}$~M$_\odot$ 
respectively. We now describe the selection of our candidate haloes.

%__________________________________________________________________
\subsection{Environment}
\label{envi}

\begin{figure}[htb]
\centering
\includegraphics[width=9cm]{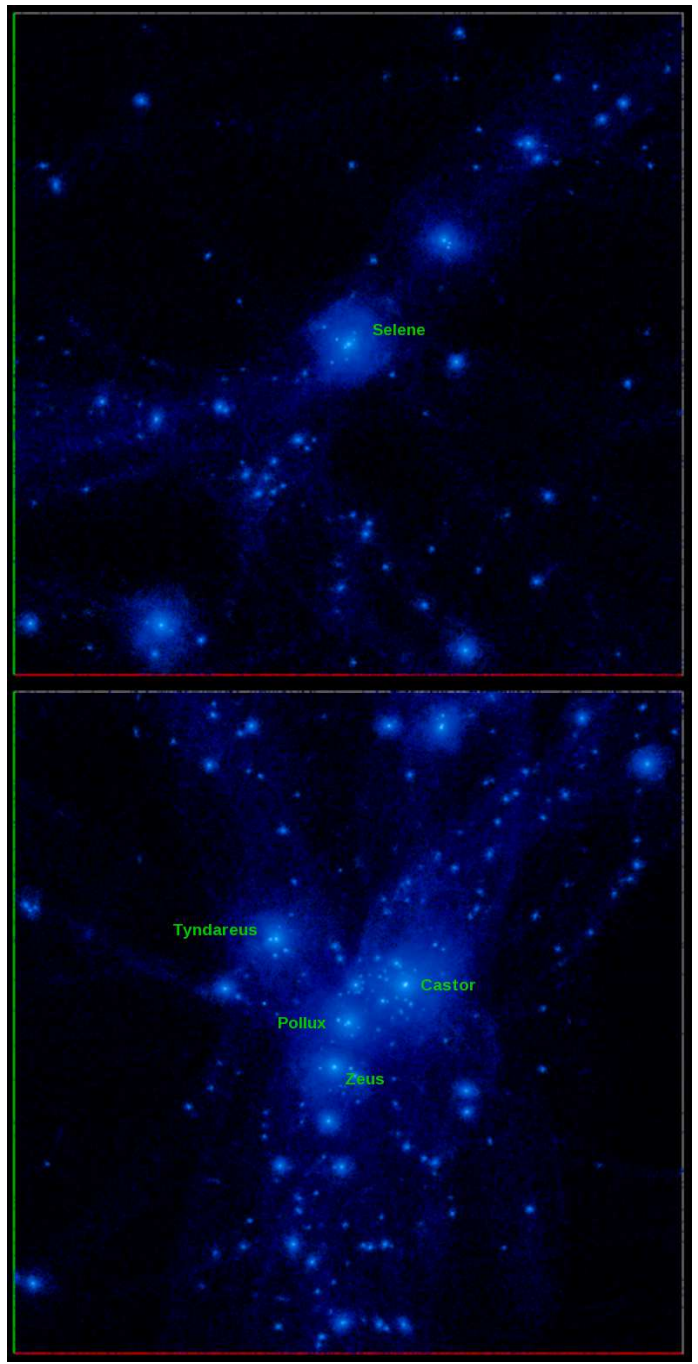}
\caption{Examples of the different environments considered in this work. The dark matter haloes of the field galaxy \emph{Selene} (top) 
and the loose group galaxies \emph{Castor}, \emph{Pollux}, \emph{Tyndareus} and \emph{Zeus} (bottom) are shown as blue particles (lighter colours correspond 
to denser regions). Images are 4$\times$4~Mpc$^{2}$ in size and have a depth of 4~Mpc.}
\label{environment}
\end{figure}

\begin{figure}
\centering
\includegraphics[width=9cm]{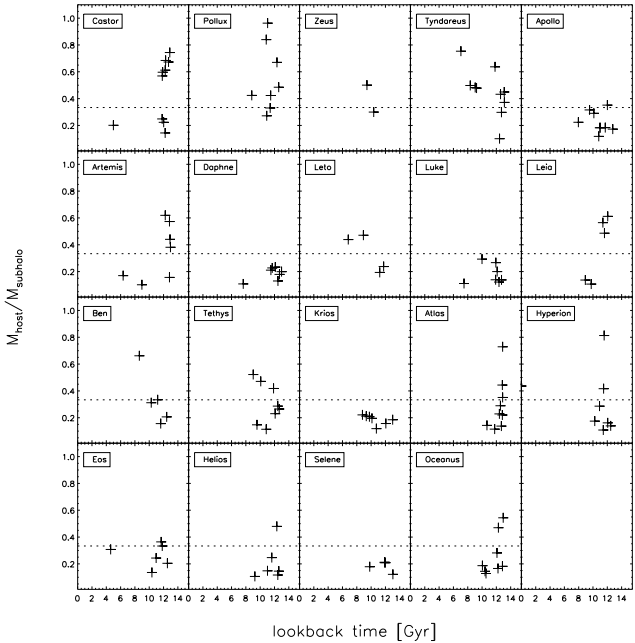}
\caption{Major mergers distributions, the top two rows are loose group galaxies and the bottom two rows are field galaxies, a 
convention that is followed throughout this work. 
The magnitude of the merger is represented by the virial mass of the host halo divided by the virial mass of the 
merging body. As stated in the text, mass ratios are calculated at the time the haloes first come into contact as the mass 
associated with merging structures is often greatly diminished by the time the disk is disrupted. The majority of mergers take 
place at early times when the galaxy halo still has a low mass and assembly is mostly hierarchical. We find that only mergers shown 
above to occur further back than 4 Gyr have any significant influence on the disk structure and that mergers can take several Gyr before they have 
descended far enough into the halo to disrupt the disk. A dashed line shows the lower bound of the commonly used definition of major mergers.}
\label{mergers}
\end{figure}

A rich literature exists comparing simulated field disk galaxies with observations of the Milky Way \citep[e.g.][]{brook04, scannapieco05, sanchezblazquez09, kobayashi11, house11, guedes11}.  
The purpose of our work here is to determine whether or not environments comparable to those of the Local Group 
- the true environment of the Milky Way - result in any measurable characteristics which would call into question 
this fundamental tenet of simulation vs observation comparison.

%-----------------------------------

The dark matter haloes of the galaxies presented here are chosen as follows. 
Dark matter haloes with virial mass (M$_{vir}$) in the range 5$\times$10$^{11}$--10$^{12}$~M$_\odot$ are considered as
loose group candidate haloes. Further selection criteria for the loose group sample are such that large groups are excluded;
each halo must have at least one companion with a comparable virial mass at a distance of 500--700~kpc but have no haloes more massive
than 5$\times$10$^{12}$~M$_\odot$ within 5~Mpc. Please note that the loose groups are not specifically constrained to be pairs,
rather they may include up to four haloes each with M$_{vir}$$>$10$^{11}$~M$_{\odot}$.

The field sample consists of galaxies that have no other dark matter haloes of mass M$_{\mathrm{vir}}$$>$3$\times$10$^{11}$~M$_{\odot}$
within a 3~Mpc radius. These field environment galaxies are far more common than the loose groups and those presented here were chosen based upon
properties such as mass, spin factor and number of mergers such that a similar range in each of these can be found in both the
loose group and field samples. The spin factors of the dark matter haloes range from 0.007 to 0.08, this range encompasses
the majority of dark matter halos and is simply used to exclude outliers. The merger trees of the galaxies are used to select loose group
and isolated haloes to avoid galaxies (in both loose group and field samples) that have any significant mergers after z=1. This is done to
ensure that disk galaxies are formed and also prevents the increased likelihood of mergers in denser environments from dominating the systematic
differences. The full sample of galaxies covers a range in virial mass from 10$^{11}$--10$^{12}$~M$_{\odot}$ and includes the field galaxies
and all the group galaxies in this mass range, i.e. massive satellites in addition to the dominant group members.

There are ten loose group galaxies (taken from three groups) and nine field galaxies. The dark matter distributions of a field galaxy and a loose group
are shown in Figure~\ref{environment}. Each sample is roughly divided into two
different resolutions corresponding to the different cosmological volumes. None of the galaxies have passed
pericenter with one another as of z=0 and so one should not expect dramatic tidal or merger effects. Two of the galaxies do have recent mergers,
but for the analysis presented here are studied at an earlier timestep (that is unaffected by the z=0 merger, as described later). The details of each galaxy can be found
in Table~\ref{tab1}. It is well known that the stellar mass fraction of simulated galaxies is too high, particularly for older stellar
populations. This is true of the galaxies presented here (stellar masses are stated in Table~\ref{tab1}) with the stellar-to-total mass fraction a factor of 2--3 times 
too high when counting all mass in the virial radius \citep{mandelbaum06, moster10, leauthaud12}. We do not believe that this has a drastic impact on this analysis. 
Firstly, the issue affects galaxies independent of environment, so comparisons of the loose group and field samples are not systematically offset by this effect. 
Secondly, the early formation of stars would lead to an overly massive spheroid. We will show later that the spheroid mass is \emph{not} extreme, 
although admittedly the disk stars are too concentrated, i.e. disk scale lengths are too short (see \S\ref{metgradients}).

A preliminary study of the metallicity gradients in these galaxies has been conducted in comparison with metallicity
gradients found in other simulated galaxies and in semi-numerical models in \cite{pilkington12} (there these galaxies are referred to as Ramses Disk 
Environment Study or RaDES galaxies); that work demonstrated that when compared with other models the RaDES galaxies tend to possess shallower 
metallicity gradients and have a more subtle evolution as a function of time, something the authors attribute to the more uniform star formation 
profile compared to other models \citep{pilkington12}. This issue is discussed further in \S\ref{metgradients}.

\begin{table*}[htb]
\begin{center}
\caption{List of the disk galaxies simulated for this work, with their total, dark matter, baryonic, stellar and gaseous masses. The spatial resolution of the simulation they originate from 
and their host environment is also quoted. The properties are calculated at z=0 except for \emph{Castor} and \emph{Eos} as described in \S\ref{merger}}\label{tab1}
%\begin{tabular}{p{1.5cm} p{2cm} p{1.5cm} p{1.5cm} p{1.5cm} p{1.5cm} p{1.5cm} p{1.5cm}}
\begin{tabular}{l c c c c c c c}
%Name & Environment & Resolution [pc] & M$_{tot}$ \newline[M$_\odot$] & M$_{DM}$ \newline[M$_\odot$] & M$_{baryon}$ \newline[M$_\odot$] & M$_{stellar}$ \newline[M$_\odot$] & M$_{gas}$ \newline[M$_\odot$]\\
Name & Environment & Resolution & M$_{\mathrm{tot}}$ & M$_{\mathrm{DM}}$ & M$_{\mathrm{baryon}}$ & M$_{\mathrm{stellar}}$ & M$_{\mathrm{gas}}$ \\
 &  & (pc) & (M$_\odot$) & (M$_\odot$) & (M$_\odot$) & (M$_\odot$) & (M$_\odot$)\\
\hline
\hline
Castor    & loose group &436. &  1.05$\times10^{12}$&  8.70$\times10^{11}$&  1.77$\times10^{11}$&  1.10$\times10^{11}$&  6.76$\times10^{10}$\\
Pollux    & loose group &436. &  4.23$\times10^{11}$&  3.48$\times10^{11}$&  7.54$\times10^{10}$&  4.99$\times10^{10}$&  2.55$\times10^{10}$\\
Zeus      & loose group &436. &  2.33$\times10^{11}$&  1.97$\times10^{11}$&  3.60$\times10^{10}$&  2.57$\times10^{10}$&  1.03$\times10^{10}$\\
Tyndareus & loose group &436. &  3.30$\times10^{11}$&  2.82$\times10^{11}$&  4.82$\times10^{10}$&  3.24$\times10^{10}$&  1.59$\times10^{10}$\\
Apollo    & loose group &523. &  8.94$\times10^{11}$&  7.39$\times10^{11}$&  1.55$\times10^{11}$&  1.06$\times10^{11}$&  4.89$\times10^{10}$\\
Artemis   & loose group &523. &  7.45$\times10^{11}$&  6.46$\times10^{11}$&  9.84$\times10^{10}$&  5.61$\times10^{10}$&  4.23$\times10^{10}$\\
Daphne    & loose group &523. &  3.09$\times10^{11}$&  2.58$\times10^{11}$&  5.14$\times10^{10}$&  2.90$\times10^{10}$&  2.24$\times10^{10}$\\
Leto      & loose group &523. &  2.49$\times10^{11}$&  2.05$\times10^{11}$&  4.38$\times10^{10}$&  2.92$\times10^{10}$&  1.46$\times10^{10}$\\
Luke      & loose group &523. &  1.13$\times10^{12}$&  9.40$\times10^{11}$&  1.88$\times10^{11}$&  1.07$\times10^{11}$&  8.13$\times10^{10}$\\
Leia      & loose group &523. &  3.93$\times10^{11}$&  3.25$\times10^{11}$&  6.76$\times10^{10}$&  4.75$\times10^{10}$&  2.01$\times10^{10}$\\
\hline
Ben       & field       &523. &  7.74$\times10^{11}$&  6.42$\times10^{11}$&  1.32$\times10^{11}$&  8.16$\times10^{10}$&  5.08$\times10^{10}$\\
Tethys    & field       &523. &  7.21$\times10^{11}$&  5.94$\times10^{11}$&  1.27$\times10^{11}$&  8.21$\times10^{10}$&  4.51$\times10^{10}$\\
Krios     & field       &523. &  5.68$\times10^{11}$&  4.78$\times10^{11}$&  8.98$\times10^{10}$&  6.15$\times10^{10}$&  2.84$\times10^{10}$\\
Atlas     & field       &523. &  6.48$\times10^{11}$&  5.55$\times10^{11}$&  9.35$\times10^{10}$&  6.15$\times10^{10}$&  3.20$\times10^{10}$\\
Hyperion  & field       &523. &  1.03$\times10^{12}$&  8.58$\times10^{11}$&  1.72$\times10^{11}$&  1.13$\times10^{11}$&  5.92$\times10^{10}$\\
Eos       & field       &436. &  4.64$\times10^{11}$&  3.92$\times10^{11}$&  7.19$\times10^{10}$&  4.68$\times10^{10}$&  2.52$\times10^{10}$\\
Helios    & field       &436. &  1.05$\times10^{12}$&  8.91$\times10^{11}$&  1.62$\times10^{11}$&  1.23$\times10^{11}$&  3.93$\times10^{10}$\\
Selene    & field       &436. &  6.07$\times10^{11}$&  5.09$\times10^{11}$&  9.87$\times10^{10}$&  6.71$\times10^{10}$&  3.16$\times10^{10}$\\
Oceanus   & field       &436. &  1.12$\times10^{12}$&  9.19$\times10^{11}$&  1.97$\times10^{11}$&  1.41$\times10^{11}$&  5.59$\times10^{10}$\\
\end{tabular}
\end{center}
\end{table*}

\begin{figure*}[htb]
\centering
\includegraphics[width=18cm]{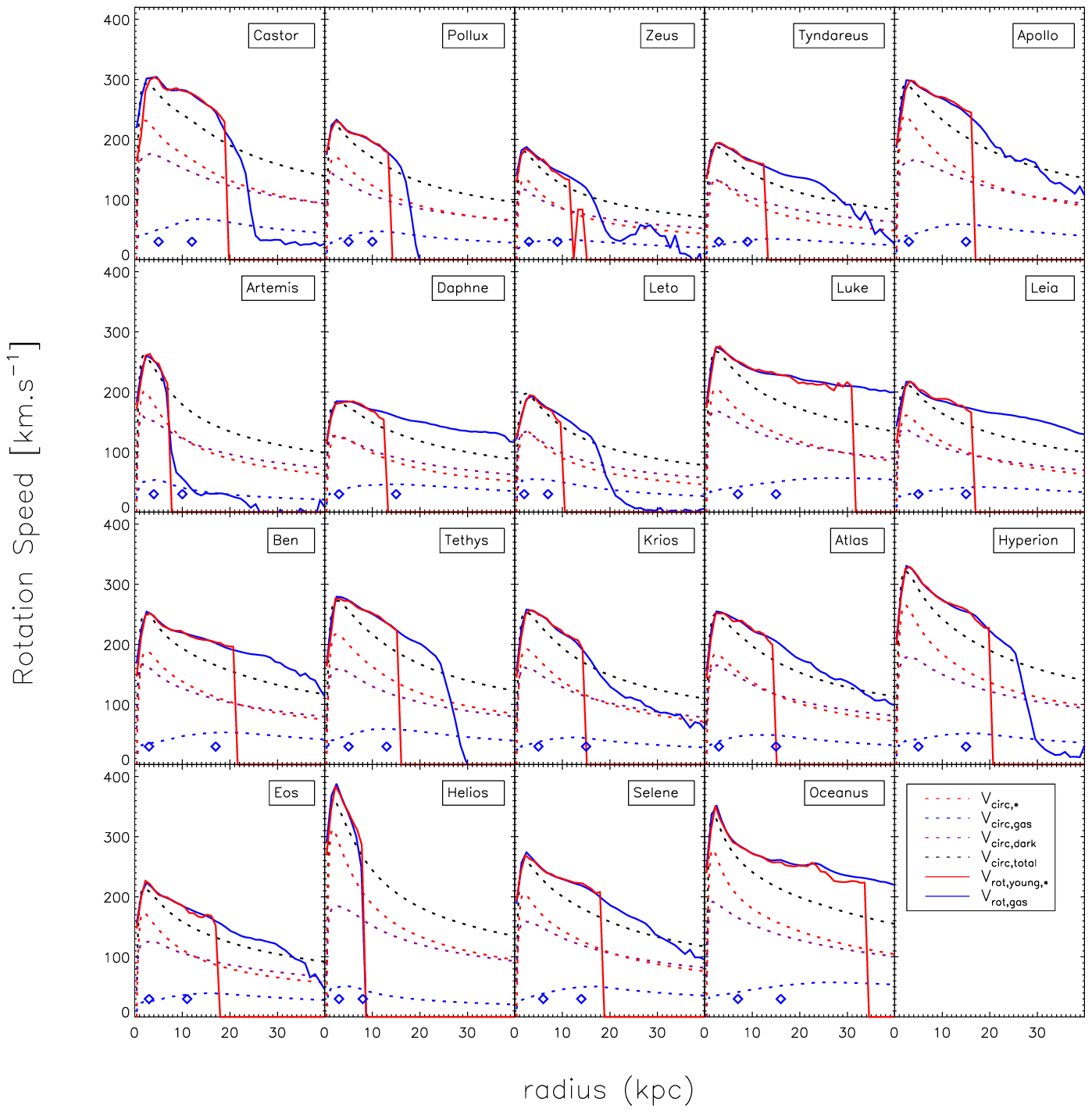}
\caption{Rotation curves showing the circular velocity for stars (red), gas (blue) and dark matter (black) as dashed lines and 
  rotation velocities (solid lines) of young stars (age$<$100 Myr at z=0) and gas as a function of radius.
  Two blue diamonds denote the inner and outer ``disk radii'' at an arbitrary vertical position. These values are chosen from inspection of the rotation curve 
  of the gas, surface density maps, stellar surface density profiles and metallicity profiles. The most conservative choice is made 
  to avoid spurious fits but to maintain consistency throughout the analysis.}
\label{rotation}
\end{figure*}

%__________________________________________________________________
\subsection{Merger History}
\label{merger}

The predominant strength of using cosmological simulations to develop this suite of
galaxies is that a self-consistent merger tree provides the basis of the evolution
for each galaxy. This ensures the conditions in the environments selected here are 
consistent with the current understanding of cosmology and not biased by artificial 
initial conditions that may affect the results. We begin by demonstrating that there 
is at least a superficial similarity between the merger histories of the two samples 
in spite of the environmental differences.

At each timestep a catalogue of haloes and subhaloes is created using the \emph{Adaptahop} 
algorithm \citep{aubert04}. The halo catalogues are then linked into merger trees for the 
selected haloes using the ``most massive substructure method'' detailed in \cite{tweed09}. 
Under this formalism, at any given branch in the tree the most massive progenitor is considered the parent halo.

It is not however a trivial exercise to define mergers in this context, and the following 
definition of a merger is adopted: a merger occurs when an object is identified as a subhalo in 
a given output but not in the previous output. In practice, the subhalo continues to exist as an 
identifiable structure orbiting the host halo for several Gyr. Due to dynamical friction, the subhalo 
gradually sinks into the potential well of the host halo and is slowly stripped of mass before dissolving 
completely, at which point all particles are attributed to the host halo. We found that in general with 
the merger time definition above there is a delay of up to 4~Gyr between the merger time and any real 
interaction with the disk, i.e., there is a delay time between the dark matter merger and the galaxy disk merger. 
The definition above is used because the amount of mass loss before the subhalo dissolves is unpredictable and the 
mass ratio of subhalo to host is a more useful quantity for the purpose of evaluating the magnitude of the
merger. The magnitude and timing of mergers for each galaxy are shown in Figure~\ref{mergers}.
This figure shows that most mergers a galaxy experiences occur at early times and that these 
early mergers tend to have mass ratios closer to unity. This is partly due to universal expansion reducing the merger 
likelihood and partly to the limited size of haloes at early times limiting the mass range and making equal mass mergers more likely.

The traditional definition of a major-merger, $M_{\mathrm{host}}/M_{\mathrm{sub}}\le3$, has been made
significantly more generous ($M_{\mathrm{host}}/M_{\mathrm{sub}}\le10$) in this work due to the low number of 3:1
mergers in the sample. The low number of such mergers is due to selection effects, these galaxies are chosen from dark matter 
haloes based partly on the merger trees and haloes with many mergers were discounted as unsuitable for hosting disk galaxies. 
A commonly used metric of the merger history is the time at which the last major-merger took place. At this point 
we note that two of the galaxies (identified by the names, \emph{Castor} and \emph{Eos}) show 
obvious signs of disturbance at z=0 due to an ongoing merger that makes the disk of the galaxy 
difficult to identify. The presence of mergers may at first be considered counter to the previously mentioned constraint that the 
galaxies have no recent major mergers, in both cases the merger did not appear in the dark matter-only simulation and only 
became apparent following the inclusion of baryons. We remedy this in both cases by analysing the galaxy disk at a timestep immediately preceding 
the disturbance. While these galaxies appears in the analysis as a low redshift (z$\simeq$0.03) late-type 
object, they may also be considered as more local irregular galaxies if the z=0 output is analysed. We conclude this section by 
stressing that the objective here is not to quantify the effect of mergers on disk galaxies but to examine 
what effect the environment might have (i.e. via ambient effects) when the increased instance of mergers is discounted. As such 
we point to Figure~\ref{mergers} as evidence that the loose group galaxies have the same diversity of merger history as do the field galaxies.

\subsection{Disk Decomposition}
\label{diskdecomp}

We now separate the stellar particles into spheroidal and disk populations using a kinematic decomposition similar 
to that of \cite{abadi03}. Details on the decomposition for 
each galaxy can be found in the appendix where the orbital circularity distributions are presented.
In short, stars are assumed to belong to either a spheroidal or a rotating disk component through analysis of the 
orbital circularity, i.e. the ratio of their angular momentum, $J_{\mathrm{z}}$ to the circular orbit angular momentum $J_{\mathrm{circ}}$ for a given particle energy. 
The successes of this method are shown in Figure~\ref{decomp}. It should be noted that
while the $J_{\mathrm{z}}/J_{\mathrm{circ}}$ distribution has two components, in the intermediate region, particles are stochastically attributed
to each component. Thus it is not necessarily true that stars identified as belonging to the spheroid did not form as disk
stars. As such the spheroidal component will contain disturbed disk stars (which is arguably
appropriate) and more critically, circularly rotating bulge stars will be attributed to the 
inner regions of the decomposed disk. This does not skew the spheroid-to-disk mass ratio 
as the vast majority of star particles have equal mass, furthermore the order of this effect should be small as shown in Figure~\ref{sfh} 
which plots the star formation history for the entire galaxy (solid line) and the disk (dashed line) and demonstrates that the selected 
disk stars well represent the stars formed at late times.

In the analysis that follows we define a disk annulus to exclude contamination by bulge 
stars and avoid halo stars at the disk edge. Figure~\ref{rotation} shows the rotation curves
for the galaxies (and circular velocities for each phase of matter) and choice of disk annulus (indicated by 
diamond symbols at an arbitrary vertical position). The outer extent of the stellar disk region is first constrained by examination of the rotation curve 
of the young stars (less than 100~Myr old), the departure of the young stars rotation curve (solid red line) from the gaseous rotation curve (solid blue line) is 
a useful indicator of the stellar disk edge. Consideration is also given to the stellar density profiles and metallicity 
gradients of each galaxy, in several cases the density profile or metallicity gradient extends beyond or falls short 
of the young disk edge. The final disk annulus is conservative to allow for gradients to be measured for each property 
over a consistent region while avoiding bias from unusual features. Despite this there are cases where the region 
over which gradients and scale-length are determined has been changed to reflect the characteristics of the galaxy in 
question. A summary of the galaxy properties can be found in Table~\ref{tab2} and star formation histories for each galaxy are 
plotted in Figure~\ref{sfh}.

%--------------------------------------------------------------------------------------------------
\subsection{Mock Observations}

\begin{figure*}
  \centering
  \includegraphics[width=14cm]{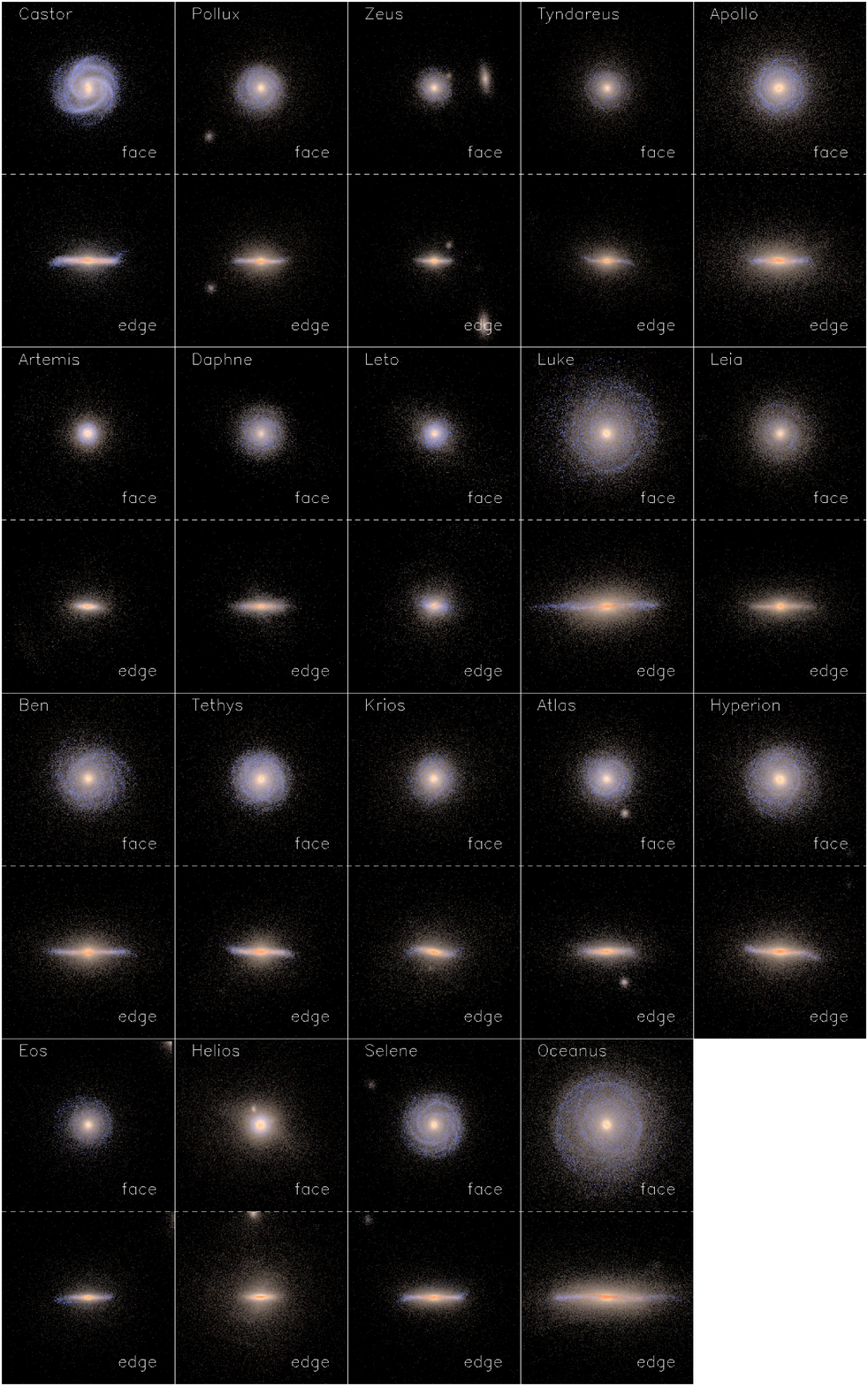}
  \caption{Mock images of the galaxies. Face- and edge-on views are separated by dashed lines, images are created 
    by combining SDSS g, r and i filters and are each 50$\times$50~kpc$^{2}$ in size. Simulation outputs at z=0 
    are used for all galaxies except \emph{Castor} and \emph{Eos} where the galaxies are shown at z$\simeq$0.03}.
  \label{images}
\end{figure*}

The galaxy's stellar and gaseous distributions, ages and metallicities have been used by the ray tracing program \textsc{Sunrise} 
\citep{jonsson06} to produce mock images. The \emph{Starburst99} stellar population models \citep{leitherer99} define the colour and 
magnitude of stellar particles. Scattering and extinction are determined by assuming that dust follows the gas phase 
metallicity distribution. Mock images of the galaxies may be found in Figure~\ref{images}, each image being 50$\times$50~kpc$^{2}$ in 
size and produced using SDSS g, r and i filters. We draw the reader's attention to the asymmetry of the more extended disks of \emph{Luke} 
and \emph{Oceanus}, the warped disks of \emph{Castor}, \emph{Tyndareus}, \emph{Krios} and \emph{Hyperion} and to the red, spheroid-dominated \emph{Helios}.

\section{Results}
\label{resu}

\subsection{Disk Fraction}

\begin{figure}[htb]
\centering
\includegraphics[width=9cm]{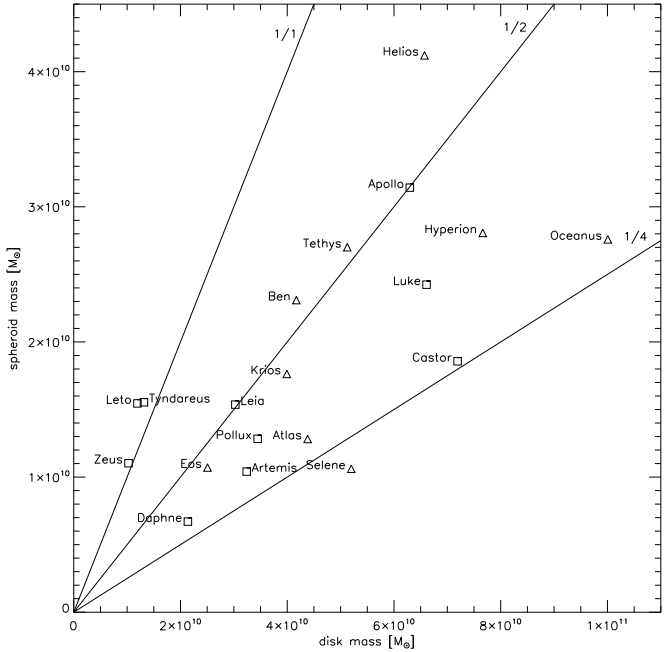}
\caption{Spheroid and disk masses as determined by kinematic decomposition of the z$\sim$0 stellar phase.
Squares and triangles indicate the loose group and field populations respectively. Radial lines have
constant bulge-disk ratios.}
\label{bulge2disk}
\end{figure}

One of the potential differences that may be seen between the galaxies in different environments is the
bulge-to-disk ratio. There is a body of evidence suggesting that a morphology-density relation exists 
\citep{dressler80, giuricin95, bamford09} due to harassment by neighbours and the increased likelihood 
of mergers. The spheroid-to-disk stellar mass ratios for the sample are shown in Figure~\ref{bulge2disk} and it is 
immediately apparent that the majority are disk dominated (spheroid/total$<$1) in spite of the spheroid containing the net mass of the halo in addition
to the bulge. Only the smallest group members are spheroid dominated (\emph{Leto}, \emph{Tyndareus} and \emph{Zeus}) with many of the larger galaxies having disk
masses exceeding spheroid masses by a factor of 2--4. There is no noticeable tendency for the galaxies with
more disturbed disks and smaller galaxies to be spheroid dominated; \emph{Artemis} has a disturbed disk and yet still exhibits
a spheroid-to-disk ratio of 0.326, likewise the low mass galaxies \emph{Daphne}, \emph{Pollux} and \emph{Eos} have quite strong disk components in
contrast with the similarly less massive (M$_{vir}$$<$4$\times10^{11}$~M$_{\odot}$) galaxies \emph{Zeus}, \emph{Tyndareus} and \emph{Leto}. Given the 
typical problem of forming too many stars, particularly at early times, in simulated galaxies one might expect the spheroid 
mass to dominate over disk mass but this does not appear to be the case; although admittedly the spheroid-to-disk mass ratios of these galaxies 
are still not as low as typical bright spiral galaxies \citep{weinzirl09}. Also note from Figure~\ref{rotation} that although the 
characteristic central peak in the rotation curve (associated with excessive concentrations of mass) is present, the decline 
of the rotation curve has more to do with the concentration of the disk. This issue affects all galaxies irrespective of 
environment and thus biases in the star formation history do not render comparison of the two environment samples unreliable.

The lack of a clear separation in the loose group and field populations in the spheroid-to-disk plot may point
to the fact that the environments here are not sufficiently different to allow the galaxies to manifest different
disk properties and that galaxies differ significantly only if they inhabit more extreme overdensities. It also reflects the dynamics 
of these groups. None of the galaxies have undergone much interaction with any other massive group member, having not passed 
pericenter with one another at z=0. This removes harassment by massive galaxies as a possible source of disruption and leaves only the 
possibility of complete mergers with smaller satellite galaxies than those shown in Figure~\ref{mergers} as a possible explanation.

%--------------------------------------------------------------------------------------------------
\begin{table*}[htb]
\begin{center}
\caption{Galaxy properties of the simulated disks at z$\sim$0: scale length, metallicity gradient (for stars with age less than 100~Myr), cold gas mass-weighted metallicity average (T$<$1.5$\times10^4$~K), stellar mass (kinematically defined disk stars), cold gas mass, and magnitudes (b and r SDSS filters). Note that for the gas mass and metallicity determinations, spatial cuts are used to exclude satellites.}\label{tab2}
%\begin{tabular}{p{1.5cm} p{1.5cm} p{1.5cm} p{1.5cm} p{1.5cm} p{2.2cm} p{2.2cm} p{1.5cm} p{1.5cm}}
\begin{tabular}{l c c c c  c c c c}
%Name & Environment & scale length [kpc] & d[Z]/dR$_{GC}$ [dex/kpc] & mean [Z] & stellar disk mass\newline [M$_\odot$] & gas disk mass\newline [M$_\odot$] & M$_B$ & M$_R$\\
Name & Environment & scale length & d[Z]~dR$_{\mathrm{GC}}^{-1}$ & mean [Z] & stellar disk mass & gas disk mass & M$_{\mathrm{b}}$ & M$_{\mathrm{r}}$\\
 & & (kpc) & (dex kpc$^{-1}$) &  & (M$_\odot$) & (M$_\odot$) &  & \\
\hline
\hline
Castor    & loose group & 3.88 & $-0.034$ & $-0.194$ & 7.19$\times10^{10}$ & 3.32$\times10^{10}$ & $-21.67 $  &  $-22.30$\\
Pollux    & loose group & 3.07 & $-0.052$ & $-0.139$ & 3.45$\times10^{10}$ & 7.64$\times10^{9}$  & $-20.58 $  &  $-21.33$\\
Zeus      & loose group & 1.76 & $-0.044$ & $-0.159$ & 1.03$\times10^{10}$ & 6.25$\times10^{9}$  & $-19.70 $  &  $-20.52$\\
Tyndareus & loose group & 2.27 & $-0.048$ & $-0.086$ & 1.32$\times10^{10}$ & 8.38$\times10^{9}$  & $-19.77 $  &  $-20.61$\\
Apollo    & loose group & 2.86 & $-0.057$ & $-0.269$ & 6.30$\times10^{10}$ & 2.25$\times10^{10}$ & $-21.22 $  &  $-22.00$\\
Artemis   & loose group & 1.75 & $-0.047$ & $-0.239$ & 3.24$\times10^{10}$ & 1.06$\times10^{10}$  &$ -20.52$   & $ -21.30$\\
Daphne    & loose group & 2.64 & $-0.060$ & $-0.139$ & 2.14$\times10^{10}$ & 1.40$\times10^{10}$ & $-20.20$   &  $-20.92$\\
Leto      & loose group & 1.56 & $-0.057$ & $-0.207$ & 1.19$\times10^{10}$ & 7.51$\times10^{9}$  & $-20.32$   &  $-21.01$\\
Luke      & loose group & 5.19 & $-0.035$ & $-0.164$ & 6.61$\times10^{10}$ & 4.58$\times10^{10}$ & $-21.30$   &  $-22.05$\\
Leia      & loose group & 3.94 & $-0.019$ & $-0.130$ & 3.03$\times10^{10}$ & 1.22$\times10^{10}$ & $-20.22$   &  $-21.06$\\
\hline
Ben       & field       & 3.85 & $-0.033$ & $-0.272$ & 4.17$\times10^{10}$ & 2.70$\times10^{10}$ & $-20.96$   &  $-21.67$\\
Tethys    & field       & 2.77 & $-0.050$ & $-0.231$ & 5.12$\times10^{10}$ & 1.52$\times10^{10}$  &$ -21.24$   & $ -21.97$\\
Krios     & field       & 2.50 & $-0.051$ & $-0.204$ & 3.99$\times10^{10}$ & 1.19$\times10^{10}$  & $-20.61$   & $ -21.43$\\
Atlas     & field       & 2.71 & $-0.042$ & $-0.170$ & 4.38$\times10^{10}$ & 1.22$\times10^{10}$  & $-20.88$   & $ -21.61$\\
Hyperion  & field       & 3.59 & $-0.040$ & $-0.199$ & 7.66$\times10^{10}$ & 1.46$\times10^{10}$  & $-21.24$   & $ -22.05$\\
Eos       & field       & 1.96 & $-0.069$ & $-0.279$ & 2.51$\times10^{10}$ & 1.14$\times10^{10}$  & $-20.07$   & $ -20.87$\\
Helios    & field       & 1.56 & $-0.037$ & $-0.069$ & 6.57$\times10^{10}$ & 5.26$\times10^{9}$  & $-21.01$   &  $-21.93$\\
Selene    & field       & 3.54 & $-0.061$ & $-0.244$ & 5.20$\times10^{10}$ & 1.66$\times10^{10}$ & $-20.83$   &  $-21.56$\\
Oceanus   & field       & 6.45 & $-0.029$ & $-0.103$ & 1.00$\times10^{11}$ & 2.99$\times10^{10}$ & $-21.61$   &  $-22.39$\\
\end{tabular}
\end{center}
\end{table*}

%--------------------------------------------------------------------------------------------------
\subsection{Metallicity Gradients}
\label{metgradients}
\begin{figure}[htb]
\centering
\includegraphics[width=9cm]{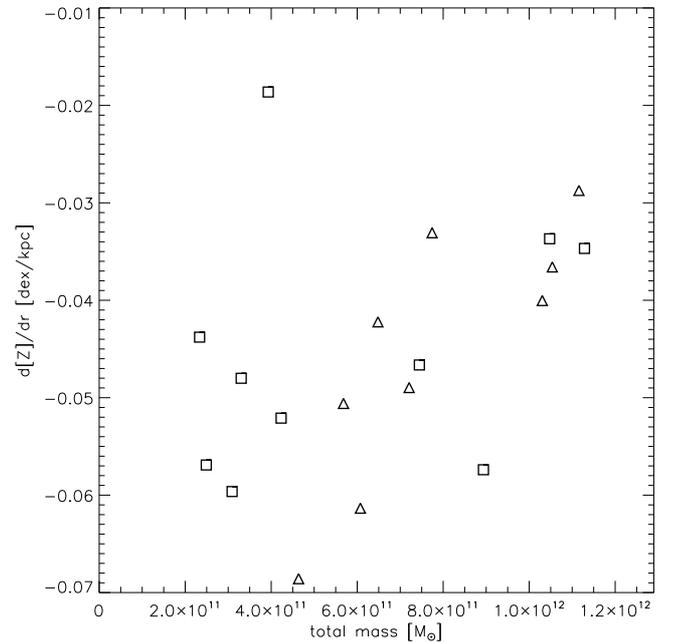}
\caption{Metallicity gradients of the disk stars at z$\sim$0 plotted against the total mass of the galaxy halo, triangles represent field galaxies, squares are 
  loose group galaxies. A trend toward shallower gradients with increasing mass is evident for both environment samples with a slight offset to flatter gradients for 
  the loose group galaxies.}
\label{absgradient}
\end{figure}

We now examine the metallicity gradients of the galaxies for evidence of environmental influences. 
Metallicity gradients of interacting and merged galaxies are known to be flatter \citep{ellison08, rupke08, kewley10, perez11} 
and as such, abundance gradients provide a probe of the dynamical mixing. To make comparisons with observed HII gradients we select 
stars that are younger than 100~Myr from the kinematic decomposition and also employing the spatial constraints described in \S\ref{diskdecomp}. 
Young stellar metallicity gradients are shown in Figure~\ref{absgradient} as a function of total galaxy mass. 
The gradients exhibited by the RaDES galaxies range from $-0.07$ to $-0.02$~dex~kpc$^{-1}$, consistent with observations by \cite{zaritsky94} 
of spiral galaxies in the field ($-0.231$ to $0.021$~dex~kpc$^{-1}$). Gradients are also calculated for spiral galaxies in \cite{vanzee98} spanning 
$-0.07$ to $-0.04$~dex~kpc$^{-1}$ and \cite{garnett97} with a range of $-0.083$ to $-0.020$~dex~kpc$^{-1}$, the RaDES galaxies are remarkably 
close to these values. 

Metallicity gradients are thought to be flatter for galaxies in denser environments and has been demonstrated to be true for close interacting 
binaries by \cite{kewley10} where HII region metallicity gradients are found between $-0.040$ to $-0.007$ dex~kpc$^{-1}$. While there is no obvious 
distinction between the two samples presented here it is worth bearing in mind that the RaDES loose group galaxies are an order of magnitude 
more distant from each other than the galaxy pairs in \cite{kewley10}. Furthermore the simulated gradients are not dramatically inconsistent 
with those measured for interacting binaries. We note that the simulated loose group galaxies do not have appreciably flatter young stellar 
gradients (with the exception of \emph{Leia}), however they likewise do not have steeper gradients, which leaves the possibility that the statistical 
sizes of the samples here may be too small to probe such a slight effect.

\begin{figure}[]
\centering
\includegraphics[width=9cm]{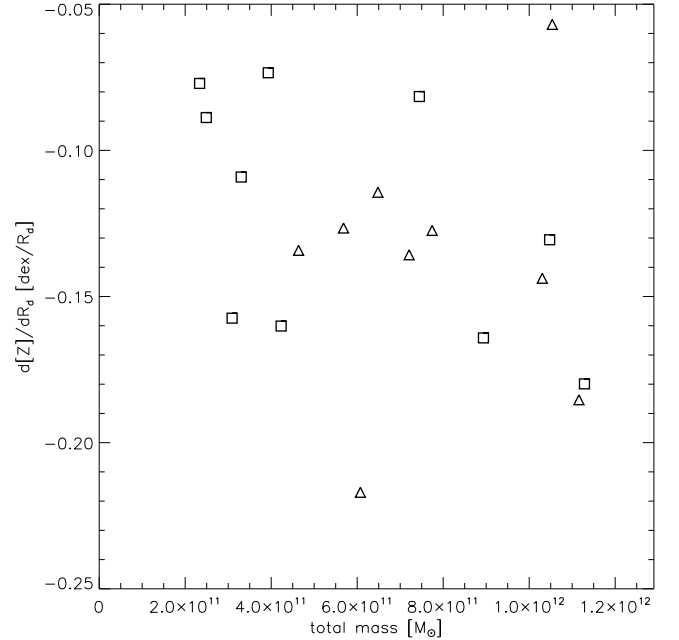}
\caption{Young stellar metallicity gradients at z$\sim$0 normalised by stellar scale length plotted against the total galaxy mass, symbols are as in Figure~\ref{absgradient}.}
\label{normgradient}
\end{figure}

Another feature of Figure~\ref{absgradient} is the trend for less massive galaxies to have steeper gradients. \emph{Leia} is a notable 
outlier from this trend, having a particularly flat metallicity distribution, however when the gradient is calculated on stars 
of all ages the trend remains and \emph{Leia} does not appear to be peculiar. This is counter to what might be expected since these 
galaxies have less massive dark matter haloes and therefore may be more easily perturbed and have flattened metallicity gradients. 
In \cite{prantzos00}, cosmologically motivated scaling relations 
are used to demonstrate that the metallicity gradients of spiral galaxies are steeper in less massive galaxies when expressed 
in {dex~kpc$^{-1}$} but not so when expressed in {dex~R$_{\mathrm{d}}^{-1}$} where R$_{\mathrm{d}}$ is the disk scale length. 
This is a consequence of the shorter scale-length of the less massive disks arising from a steeper star formation rate profile 
which results in a greater metal production rate in the inner disk compared with the disk periphery. This behaviour is supported 
by observations \citep{garnett97, vanzee98} where it is shown that less luminous spiral galaxies have steeper gradients than 
brighter galaxies when the absolute gradient is measured. When observed gradients are normalised to the disk scale-length however 
no significant variation with luminosity is apparent. Another finding of \cite{garnett97} is that the dispersion of absolute gradients 
is larger for less luminous galaxies but when normalised the dispersion is consistent with brighter galaxies pointing to the existence 
of some degree of co-evolution of metallicity and density profiles. The scale length normalised metallicity gradients of young stars 
are shown in Figure~\ref{normgradient} and confirm that metallicity gradients likely have a common origin with the stellar density 
profile. This is explored in more depth in \cite{pilkington12} where clear links are made between the star formation profile and 
metallicity gradients. The absolute metallicity gradients of the RaDES galaxies are consistent with values found in 
literature \citep{garnett97} however the normalised gradients are around an order of magnitude flatter suggesting that the 
measured density profiles are too steep, a feature that is consistent with the known issue of excess star formation at early 
times and peaked rotation curves in simulations \citep{navarro91, governato04, guo10, sawala11}. A result that is relevant 
to this work is that of \cite{dutil99}. Here the authors find that HII metallicity 
gradients are flatter for earlier morphological types, but critically, that the trend is weaker when the gradients are normalised to 
some isophotal or effective radius.  While we do not show the correlation between gradient and morphology (indeed no attempt is made 
to identify the classical morphology of these galaxies) it is clear that the metal gradient has some degree of co-evolution with the scale length.

%this graph needs to be changed

\subsection{Colour-Magnitude Diagram}

\begin{figure}[htb]
\centering
\includegraphics[width=9cm]{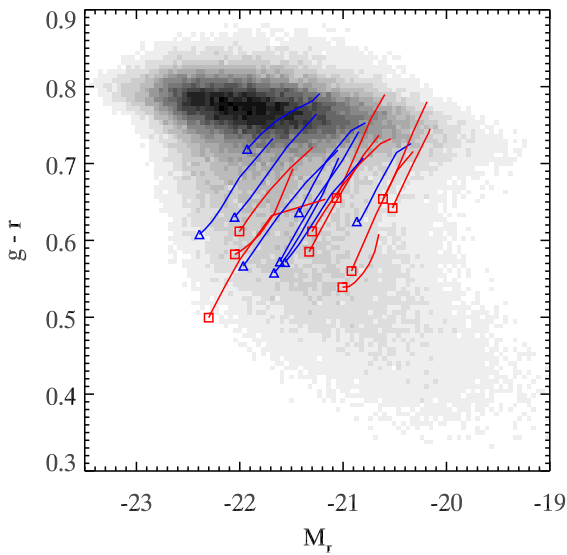}
\caption{Colour-magnitude diagram. The background data are SDSS galaxies with no inclination correction. Symbols follow the paper 
  convention and denote the face-on values. The tails traces the change in orientation from face- to edge-on.}
\label{cmd}
\end{figure}

We use \textsc{Sunrise} to produce seven different projections ranging from face-on to edge-on and display the values in a colour-magnitude 
diagram (Figure~\ref{cmd}) that overplots the observed (uncorrected) colour-magnitude diagram from SDSS data \citep{bailin08}. We represent the change in magnitude and colour as a function of projection 
angle for each galaxy in Figure~\ref{cmd} with an arc that starts with a symbol denoting the face-on projection (symbol type follows paper conventions), the end of the arc 
denotes an edge on projection. Face-on magnitudes for each galaxy are given in Table~\ref{tab2}. 
Almost all the galaxies populate the blue cloud with only \emph{Helios} appearing within the red sequence. Much as expected, as the galaxies becomes more 
inclined they appear to dim and redden; many of the galaxies therefore have edge-on projections that stray into the dimmest end of the red sequence. 
There is a selection effect at work here as the galaxies are \emph{a priori} chosen to be disk galaxies with ongoing star formation and that only \emph{Helios} is particularly red is 
reassuring. The colour of the galaxies reflects that while the disk stars are too concentrated there is no critical over production of stars at high redshift which 
would bias the galaxies to the red regions of Figure~\ref{cmd}.

There is no apparent separation of the two samples once the obvious outlier of \emph{Helios} has been discounted. There is some observational evidence that the 
colour distribution of late-type galaxies is only weakly dependent on environment and that it is more strongly influenced by the luminosity or mass through 
intrinsic evolution \citep{balogh04b}. The distribution in r-band magnitude is consistent with the mass of each galaxy and the colour of galaxies 
can be considered a probe of their star formation history (Figure~\ref{sfh}). Taking \emph{Helios} as an example, we see an initially prolonged star formation phase in comparison with 
other galaxies that have star formation rates that are less disparate throughout time. We examine the impulsive interactions of the disk with respect to 
kinematics in the next subsection.

%--------------------------------------------------------------------------------------------------
\subsection{Sigma-Age Relation}
\label{sigagesec}

\begin{figure*}[htb]
\centering
\includegraphics[width=18cm]{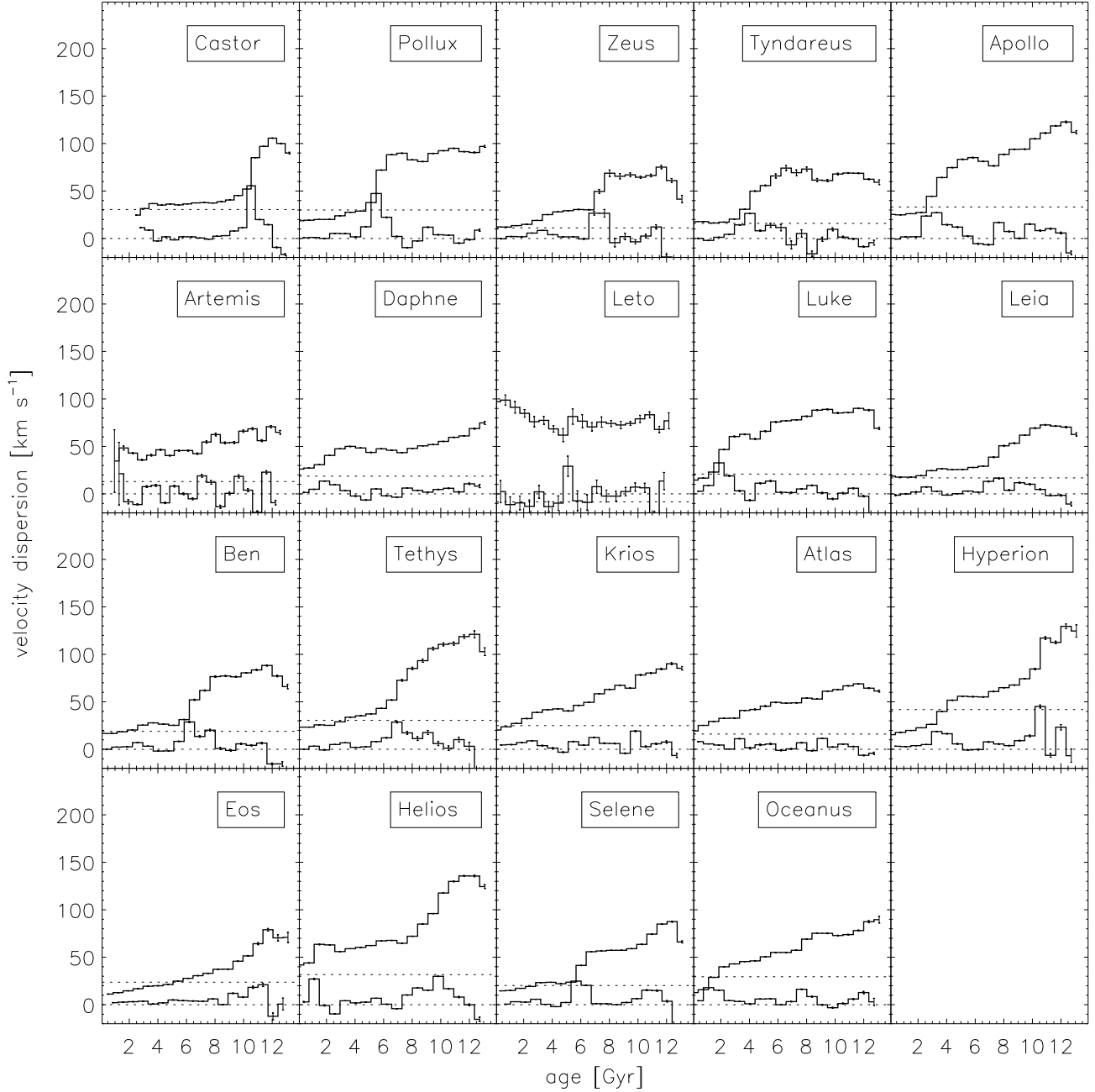}
\caption{Present day (z=0 in most cases, see \S\ref{merger}) stellar velocity dispersion as a function of age for an annulus of thickness 6~kpc centred on $2 \times R_{\mathrm{d}}$ and 
a height of 3~kpc above and below the equatorial plane. The lower line is the age-derivative of this function, 
$\mathrm{d}\sigma/\mathrm{d}age$. Horizontal dotted lines define the zero point and a step-threshold that is $5\times \langle \mathrm{d}\sigma/\mathrm{d}age \rangle$.}
\label{sigage}
\end{figure*}

We now move away from morphology and chemistry to examine kinematics and use the temporal behaviour of the stellar velocity dispersion 
($\sigma$) for this purpose. The velocity dispersion of a region analogous to the ``solar neighbourhood'' is used to quantify the 
influence that external interactions have on the kinematics of the disk. To remove the bias that arises from the velocity dispersion 
gradient as a function of disk radius we select stars from an annulus of thickness 2~kpc, centred on 3 times the disk scale length 
and a height of less than 3~kpc above and below the equatorial plane. Figure~\ref{sigage} shows the velocity dispersion of stars, 
$\sigma$, as a function of their age at z=0. While many of the RaDES galaxies attain the observed velocity dispersion of the Milky Way 
(10--20~km~s$^{-1}$ found by \cite{soubiran05, soubiran08, holmberg07}) when the youngest stars are considered, the older populations 
have far greater dispersions than observed. The $\sigma$-$age$ dispersion relations shown in Figure~\ref{sigage} exhibit a greater 
increase in dispersion as a function of age and the appearance of more discrete steps than are apparent in observations. Greater velocity 
dispersions are to be expected in simulations as less than ten resolution elements are found in the would-be thin-disk (if such a 
structure were resolvable). The high early velocity dispersions found even in field galaxies with the fewest mergers (e.g. \emph{Krios} 
or \emph{Selene}) may indicate that these galaxies experience excess kinematic heating from mergers. Another possibility is that very 
early in the galaxies formation cold flows (rather than discrete mergers) are responsible for forming the majority of stars. As it is well 
known that cosmologically simulated galaxies form too many stars at early times (see Table~\ref{tab1} for the stellar mass fractions) it 
is possible that the high velocity dispersion of old stars is simply reflecting this shortcoming. We note however that maps of the gas 
distribution at high redshift clearly show discrete gaseous objects merging and no evidence is found for cold streams after z=2.

The $\sigma$-$age$ relations do not at first display any obvious distinction between field and loose group environments, 
so a more rigorous analysis is called for. We have attempted to quantify the difference between a ``stepped'' and a 
smooth $\sigma$-$age$ relation by looking for spikes in the age-derivative of $\sigma(age)$. If the age-derivative of $\sigma(age)$ 
exhibits spikes above some significance level (the step-threshold) it will betray the existence of steps in the velocity dispersion. Galaxies that 
have steps with magnitude exceeding a threshold (the step-threshold) that is a factor of $\beta$ greater than the average of the 
age-derivative of $\sigma$ are defined as having a ``stepped profile''. The value of $\beta$ is chosen in the range 4--7 as values that 
are too low will not discriminate small variations in the age-derivative from larger steps and a value that is too large will classify 
all steps as being normal. This test is repeated for different spatial selections in each galaxy (annuli with 2-4$\times$$R_d$ in radius 
with a width of $\pm$2-4~kpc including all stars within 3~kpc of the disk plane) to avoid bias that may arise from sampling a spiral arm 
and for different values of $beta$ in the range 4--7. In each case the field sample has approximately half the number of galaxies with step 
features than does the loose group sample. An example is shown in Figure~\ref{sigage} for $\beta=5$ (for stars in an annulus of thickness 
6~kpc centred on $2 \times R_{\mathrm{d}}$ and a height of 3~kpc above and below the equatorial plane) in which seven out of ten 
loose group galaxies have stepped profiles while only three of the nine field galaxies do. The binning in age usually places at least 
several hundred stars in each age bin making the uncertainty in the velocity dispersion (and propagated uncertainty in the discrete age-derivative) 
very small, these are plotted as 2 sigma error bars in Figure~\ref{sigage}. In the extreme case of taking $\beta=8$ none of the field galaxies have 
detected steps while four of the loose group galaxies do.

While each choice of annulus and $\beta$ results in the loose group sample having a greater fraction of galaxies with stepped profiles than the field 
sample, the number of galaxies studied here is not enough for this result to be conclusive and a more complete sample would be required to make it so. 
Nevertheless we now discuss the implications of the trend if it is real. The physical mechanism shaping the $\sigma$-$age$ relation can be simply 
thought of as stars being heated to greater dispersion by all mergers subsequent to their formation and to an extent that depends on the severity of 
the merger \citep{villalobos08, dimatteo11}. A series of gentle mergers therefore results in a smoother decline in dispersion towards younger stars 
\citep{kazantzidis08}  while a large and disruptive merger excites all stars formed previously to a plateau that gives a more step-like appearance 
to the $\sigma$-$age$ relation \citep{brook04}. This contrasts with the conclusions of \cite{house11} where stars are found to form with a velocity 
dispersion and retain it as a signature of the gas state at that time, however in either scenario the analysis performed in this work is a valid measure 
of the effect of disk disruption on stellar dynamics. The results shown in this work tentatively suggest that in spite of a superficially similar 
major-merger history, there may be some differences in the mergers experienced by the galaxy disk or the interaction histories. Firstly galaxy disk major 
mergers could impact the galaxies differentially depending on the gas fraction \citep{cox06, hopkins09, lotz10}, the orbital configuration 
\citep{barnes02, robertson06} and the large-scale tidal field \citep{martig08}. Secondly, the number of minor mergers or interactions with orbiting satellites 
could have an impact in shaping the $\sigma$-$age$ relation \citep{quinn93, abadi03, bournaud05}. Third, the interaction of the galaxy with the intragroup medium may have 
an impact as found by \cite{bekki11} wherein the authors find that repetitive harassment in groups can lead to star formation bursts and disk heating.

%__________________________________________________________________
\section{Conclusions}
\label{disc}

We present a suite of cosmological simulations with the intention of comparing 
field galaxies with galaxies in Local Group environments. The galaxies are taken 
from cosmological simulations where a zoom method is used to allow sub-kpc resolution 
while simultaneously accounting for large scale structure formation. A kinematic 
decomposition has been performed to separate disk stars from spheroid stars and we have 
analysed the morphology of the galaxies. While all galaxies studied here have larger stellar 
mass fractions than observations dictate they should, this issue is uniform and does not 
affect comparisons between the two environments. We have also examined the metallicity gradients 
finding trends with mass but a very weak or non-existent correlation with environment. 
Finally the stellar velocity dispersion is studied and evidence of a dependence on environment 
is found in the signature of impulsive heating in group galaxies. The results of this work 
are pertinent to the comparison of simulated field galaxies with observations of the Milky Way. 
The conclusions of this work are summarised here:

\begin{enumerate}
\item No distinction between loose group and field galaxies is seen when considering 
the spheroid-to-disk ratio although examination of galaxies with greater spheroidal components shows 
that they all have interactions that disturb their disk rather than forming from kinematically hotter 
gas. However this is far from conclusive as there are only four galaxies with significantly higher spheroidal 
components of the total 19.

\item Metallicity gradients of loose group galaxies are very similar to those of field galaxies 
with the same disk mass, a result that is still consistent with observations of strongly interacting galaxies \citep{kewley10} though 
non conclusive evidence is seen that loose group galaxies should have significantly flatter gradients compared with their counterparts 
in the field. The absolute gradients are consistent with observations \citep{zaritsky94, garnett97, vanzee98} yet when normalised by disk scale-length 
gradients are an order of magnitude flatter than observed suggesting that density profiles are too concentrated. Observations also show that 
more massive spiral galaxies have flatter gradients, this has previously been matched by semi-analytical models \citep{prantzos00} using scaling relations 
but the trend has now also been shown to exist for our numerically simulated galaxies. We also find a link between metallicity gradients 
and stellar density gradients that suggests that galaxies in the mass range studied here have similar metal gradients 
when expressed in dex/R$_{\mathrm{d}}$ and that variance in this value may be attributable to radial migration or disruptive mergers.

\item Examination of the age-velocity dispersion relation reveals that as expected the velocity dispersion of old stars in the 
simulated galaxies is greater than observed for the Milky Way disk. Loose group galaxies exhibit more \emph{stepped} relations 
that suggest mergers/harassment do have a greater impact on the loose group galaxies than field galaxies. This is at odds with 
the apparent similarities in the major merger frequency of the two samples and suggests that the major merger history of dark matter haloes 
may not be an accurate probe of the galaxy disk merger history. We note however that the relatively small sample size means that 
it is not conclusive whether or not this is a real effect depending on environment or simply because the loose group galaxies 
happen to have more turbulent formation histories independent of environment; certainly the individual formation history of each 
galaxy will impact on its velocity dispersion evolution and this would be consistent with the other findings of this work.
\end{enumerate}

The main conclusion to come from this work is that in such sparse environments where the galaxies are not directly interacting 
galaxies exhibit different properties depending on individual merger histories and infall rates but that loose groups environments 
are only very weakly different to the field.  It has been suggested for cluster galaxies that it is likely that galaxies are shaped 
more by direct mergers and their own secular behaviour rather than the large-scale environment that impacts the aforementioned only 
indirectly \citep{mcgee08}. Structures further than $\sim$1~Mpc distant have little influence for galaxies with masses presented here.

The method used to define mergers in this work may be insufficient to link mergers to the signatures of the impact they have on the disk 
properties and a future study to follow this should develop a larger suite using only dark matter simulations to quantify the satellite 
distribution and minor merger rates with a greater statistical significance. Simulations at higher resolution should also be employed to 
determine conclusively if any systematic difference in metallicity gradients exists. We finish by stating that at the resolutions considered 
here simulated galaxies may be safely compared with Milky Way properties whether they inhabit loose group or field environments, however 
attention must be given to the aggregated merger properties, mass and internal structure for such comparisons to be meaningful.
%__________________________________________________________________
\begin{acknowledgements}
The authors would like to acknowledge Romain Teyssier for both access 
to \textsc{ramses} and for helpful discussions regarding its use. We are 
grateful to referee Sebastien Peirani for the many suggestions that have greatly improved 
this manuscript. CGF acknowledges 
the support of STFC through its PhD Studentship programme (ST/F007701/1). BKG, CBB, 
SC, LMD, acknowledge the support of the UK's Science \& Technology Facilities Council 
(ST/F002432/1 \& ST/H00260X/1). BKG acknowledges the generous visitor support provided 
by Saint Mary's and Monash Universities. We also acknowledge the UK's National Cosmology 
Supercomputer (COSMOS), and the University of Central Lancashire's High Performance Computing Facility.
SC acknowledges support from the BINGO Project (ANR-08-BLAN-0316-01).
LMD acknowledges support from the Lyon Institute of Origins under grant ANR-10-LABX-66.
\end{acknowledgements}

\bibliographystyle{aa} 
\bibliography{rades}

\onecolumn
\appendix
\section{Additional Galaxy Properties}

Here we describe some properties of the galaxies that are not directly discussed in the main paper body but are still 
useful as supplementary material. We begin with some general comments on particular features of each galaxy that are of interest.

\begin{itemize}
\item \emph{Castor} is the only galaxy in the sample to exhibit a bar, perhaps reflecting the greater resolution or the more isotropic 
nature of its group (compared with, for example, the filamentary structure of the \emph{Apollo} group). It also presents the clearest 
example of spiral structure of the entire sample. The spiral structure presents challenges when quantifying the stellar scale 
length as the arms present a bump in the density profile. The young stars present in the arms mean that this is even more 
pronounced when measuring the brightness profile. \emph{Castor} also has the most pronounced (and abrupt) disk warp, initially this was 
believed to be evidence of poor resolution at the disk edge but investigation has revealed no particularly favoured alignment of 
the disk warps in this sample. Analysis of \emph{Castor} has been conducted on a snapshot slightly before z=0 due to the irregular shape 
induced by a late merger in this snapshot.

\item \emph{Artemis} is unusual in that it has a relatively massive dark matter halo (7.45$\times10^{11}$~M$_\odot$), reasonably 
low spheroid-to-disk fraction (0.32) and flat metallicity profile (-0.0068 dex/kpc), yet its disk scale length is only 1.87~kpc and is truncated 
at a radius of 7~kpc. There exists a gaseous polar disk (not dense enough to form stars) and yet the vertical velocity 
dispersion changes very little as a function of age. This suggests that the last major merger experienced by \emph{Artemis} left star-forming 
gas with a similar velocity dispersion to the older stars, an effect not seen in the other galaxies.

\item \emph{Leto} is the least massive galaxy within the \emph{Apollo} group and the most spheroid dominated of the galaxies. This spheroid is not composed 
of older stars as in the other galaxies, there is a significant fraction of the spheroid stars that are young. This is the result of a 
low star formation rate at early times compounded by a recent, disruptive event that is evident in the velocity dispersion-age relation.

\item \emph{Eos} undergoes a merger at late times that leaves it with an extremely irregular morphology at the last time step, analysis of this galaxy 
is performed on a snapshot prior to this event.

\item \emph{Helios} is the most early type galaxy of all. Despite its great mass it is the reddest galaxy and has young stars with around twice 
the vertical velocity dispersion of much of the rest of the sample and a prolonged early star formation episode, this contrasts with the lack 
of an identifiable late merger to result in such a morphology.

\item \emph{Selene} has few mergers in its history and is one of the most quiescent galaxies, forming the largest disk fraction of all the galaxies and presenting 
definite spiral structure.

\item \emph{Oceanus} has the greatest stellar mass in the sample (though it is among others with comparable halo masses) and has a rotating gaseous disk that extends 
as far as 40~kpc from the centre. This disk is dense enough to form stars and hence this galaxy has an extremely long scale length (6.63~kpc) and one of 
the flattest metallicity gradients. 
\end{itemize}

\begin{figure}[htb]
\centering
\includegraphics[width=18cm]{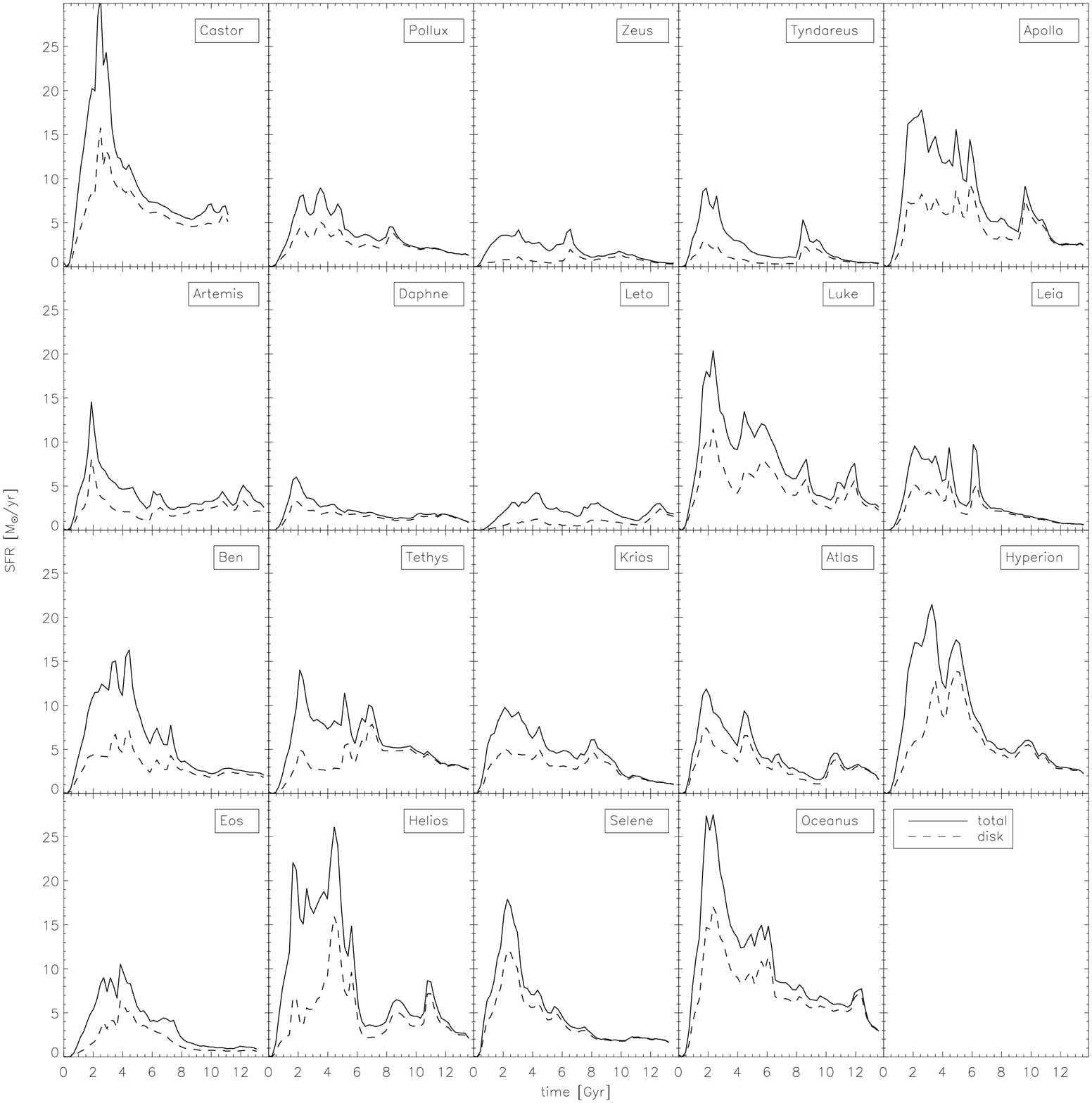}
\caption{Star formation histories of the sample galaxies. The star formation rate of all stars within the virial 
radius at z=0 are shown by the solid line, the dashed line is all stars tagged as disk stars at z=0.}
\label{sfh}
\end{figure}

One of the key ways of understanding the formation of galaxies is by examining the star formation history of the different components. The distribution of
star formation in time tells us a great deal about how the different components of a galaxy form. In the course of this work the signature of mergers were found 
to be identifiable in the bursts of star formation seen in Figure~\ref{sfh}. These bursts can in some cases be associated with steps in the velocity dispersion 
described in \S~\ref{sigagesec}, however we find that the magnitude of the star formation bursts is a poor indicator of the strength of the kinematic disturbance 
induced by the merger from which they both originate. The dichotomy of the star formation burst and the kinematic effects of mergers is yet more evidence that the 
signature of a merger depends on the gas fraction or phase space configuration of the merging bodies. Note the restrained recent star formation of the disrupted 
galaxy \emph{Artemis} compared with the more disk dominated \emph{Apollo} or \emph{Oceanus}.\\

\begin{figure}[!ht]
\centering
\includegraphics[width=18cm]{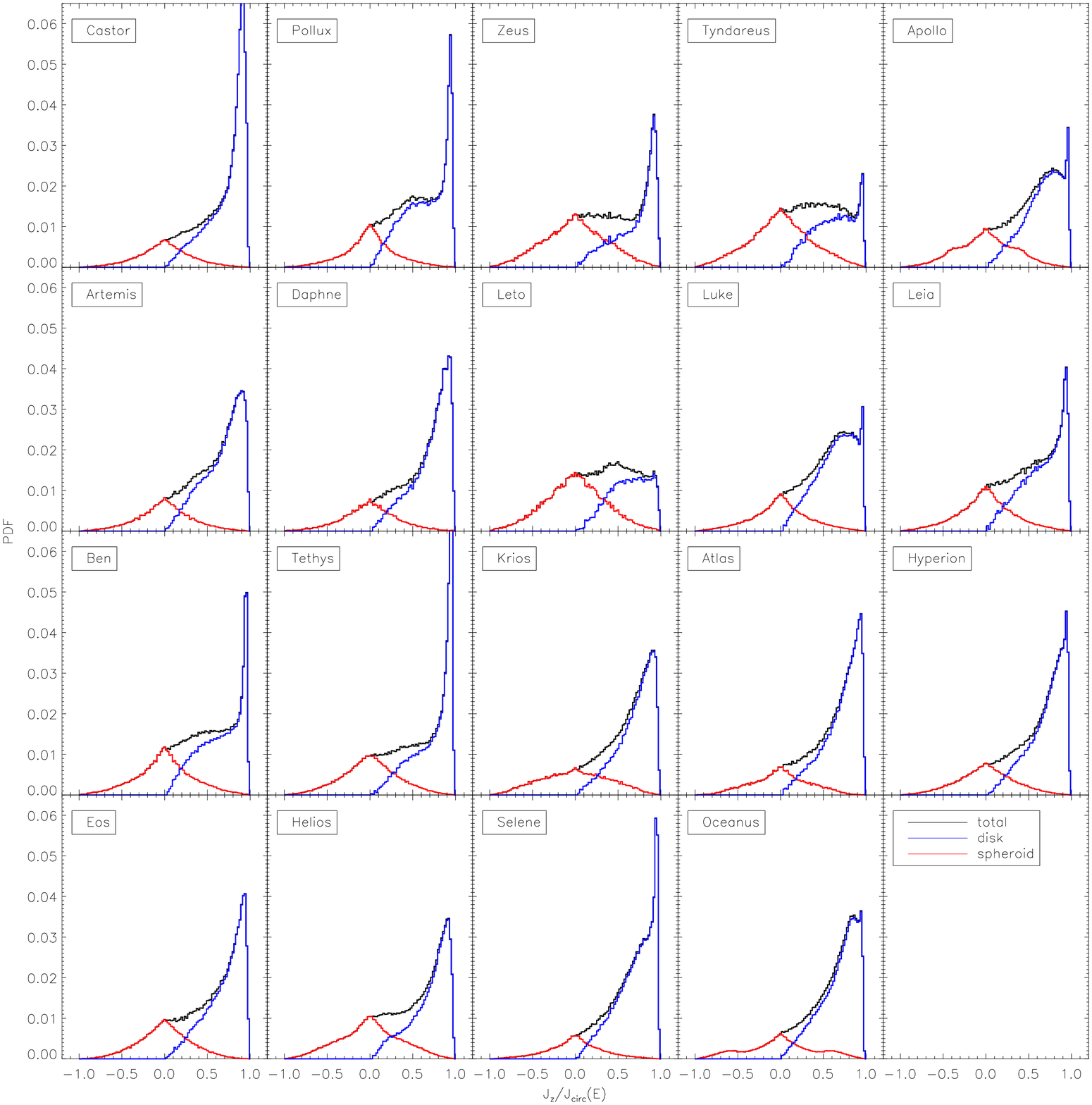}
\caption{$J_{\mathrm{z}}/J_{\mathrm{circ}}$ distribution of stars within the virial radius (black). Blue and red lines shows the 
distribution of the disk and spheroid components respectively. Note the existence of a third intermediate component 
included in the distribution that is associated with the disk in some of the galaxies.}
\label{decomp}
\end{figure}

For the analysis of the simulated disk galaxies presented in this work to be as uniform as possible we performed a kinematic decomposition to examine 
the disk stars with minimal contamination from halo and bulge stars. The decomposition employed follows the \citet{abadi03} method of using the angular momentum 
of stars compared with the angular momentum expected for rotating stars. The distribution of the relative angular momentum is shown in Figure~\ref{decomp} with 
blue and red lines highlighting the distribution on disk and spheroid stars respectively.

\end{document}